\documentstyle[12pt,epsfig]{article}
\def\comma{ \; , }
\def\period{ \; . }
\def\Del{ \nabla }
\def\di{ \partial }
\def\dir{ \partial_r }
\def\minus{ \mbox - \, }

\def\RdBk#1{ \left(\,#1\,\right) }
\def\SqBk#1{ \left[\,#1\,\right] }

\begin{document}
\title{\large\bf Topological Black Holes -- Outside Looking In}
\author{R.B. Mann$^a$  \\
Department of Physics \\
University of Waterloo, Waterloo, Ontario, Canada N2L 3G1\\ 
$^a${\it mann@avatar.uwaterloo.ca} \\ }
\date{\today \\ WATPHYS TH-97/12}
\maketitle
\begin{abstract}
I describe the general mathematical construction and physical picture of
topological black holes, which are black holes whose event horizons are
surfaces of non-trivial topology.  The construction is carried out in
an arbitrary number of dimensions, and includes all known special
cases which have appeared before in the literature. I describe the
basic features of massive charged topological black holes in $(3+1)$
dimensions, from both an exterior and interior point of view. 
To investigate their interiors, it is necessary
to understand the radiative falloff behaviour of a given massless
field at late times in the background of a topological black hole.
I describe the results of a numerical investigation of such behaviour
for a conformally coupled scalar field. Significant differences emerge
between spherical and higher genus topologies.
\end{abstract}

\pagebreak
\section{Introduction}
\label{man-intro}

It has become clear over the last decade that black holes have an essential
role to play in the development of a quantum theory of gravity.  Each aspect
of the physics of black holes -- their formation due to gravitational collapse, 
their interior structure after formation, their thermodynamic properties, the 
singularities cloaked by their event horizons, and the endpoint of their
evolution after thermally evaporating -- present a set of interconnected
puzzles whose ultimate resolution presumably entails the development of a 
full theory of quantum gravity.  
The most promising developments along this line have been in terms of the
string-theoretic and Ashtekar proposals for quantum gravity, although a
number of other candidate ideas exist, including non-commutative geometries,
gauge-theoretic formulations, quantization of topologies,
gravity as an induced phenonemon, and so on.  These ideas are not mutually
exclusive, and it is conceivable that the full theory of quantum gravity 
could contain some elements of each.  Despite enormous effort, however,
a full theory of quantum gravity remains elusive.

However the efforts expended towards achieving such a theory have yielded a
large and still growing panoply of black objects, including black holes, 
black strings, and black branes.  In general a spacetime containing a black 
object is one which has at least one event horizon: a surface which bounds
a region of spacetime that is causally cut off from future timelike infinity.
This surface is a null surface beyond which light (and hence all other forms
of mass-energy) cannot escape.  Each black object presents us with a
theoretical laboratory in which we can test some of our basic ideas about
the fundamental nature of gravity and its interactions with other forms of
mass-energy.  

One of the more interesting black objects to have entered the scene in recent
times arose from the realization that certain identifications in 
anti-de\thinspace Sitter spacetime in $(2+1)$ dimensions yields a spacetime which can 
be interpreted as a black hole of definite mass $M$ and angular momentum $J$
in a spacetime with cosmological constant $\Lambda = -1/\ell^2$.
The construction was originally given by Ba{\~n}ados, 
Teitelboim and Zanelli \cite{man-btz} and is generally referred to as the BTZ black hole. 
Although originally presented as a formal, mathematical construction 
\cite{man-btz,man-bhtz}, it was quickly realized that these black holes 
can `physically' form from the collapse of $(2+1)$-dimensional matter \cite{man-ros}
and indeed have many features in common with their more conventional 
$(3+1)$-dimensional counterparts \cite{man-revs}. 

Since the exterior spacetime of a collapsing $(2+1)$ dimensional black hole is locally
anti-de\thinspace Sitter, its singularity structure differs considerably from
that of its Schwarzchild anti-de\thinspace Sitter counterpart in $(3+1)$
dimensions. The Kretschmann scalar 
${\cal K} = R_{\mu\nu\alpha\beta}R^{\mu\nu\alpha\beta}$ has a power-law divergence 
at the origin in the latter case, whereas for the BTZ black hole it is 
everywhere finite. Instead the BTZ black hole has a causal singularity, which occurs
where the identifications surfaces merge in $(2+1)$ dimensional anti-de\thinspace Sitter
spacetime. These properties suggest that a study of its interior structure
along the lines of that carried out for mass inflation of
$(3+1)$ dimensional Reissner-Nordstrom-Vaidja
black holes \cite{man-PI} would be interesting. Such a study was carried out by
Chan {\it et. al.} for the rotating BTZ-Vaidya black hole \cite{man-chan} and yielded
several interesting results.   Assuming the radiative falloff in the BTZ-Vaidya
solution has a power-law tail, the mass function diverges at the Cauchy horizon in
the interior of the black hole, but all the scalar curvature invariants remain finite,
and tidal distortions remain bounded. However some higher-derivative curvature scalars 
diverge. This forms an interesting test \cite{man-wang} of the Konkowski-Helliwell 
conjecture, which predicts the stability of Cauchy horizons based upon the 
behavior of test fields, and in the case of instability also predicts the nature of 
the singularities produced \cite{man-konk}.   More recently, it has been shown
that the assumption of a power-law falloff for the BTZ-Vaidya black hole is not 
correct, and that the falloff rate is exponential \cite{man-jchan}.  This does not 
change any of the basic results mentioned above provided the falloff rate is 
sufficiently weak.  However for $|\Lambda| J^2/M^2 > .64$, the
exponential falloff rate is so large that mass inflation is cut off.

Intriguing as these effects are, their restriction to $(2+1)$ dimensions suggests
at least that a considerable amount of caution be exercised in applying them to
$(3+1)$ dimensions.  One way of extending these results to $(3+1)$ dimensions is to
embed the BTZ solution within a $(3+1)$-dimensional dilaton theory of gravity. The
resultant solution may be interpreted as a spinning black string \cite{man-kal}
and it exhibits mass inflation effects similar to its $(2+1)$ dimensional BTZ
counterpart \cite{man-bstr}.

However within the last year it has been shown that higher-dimensional 
generalizations of the $(2+1)$-dimensional BTZ construction exist.  This
yields an interesting new set of black holes with event horizons of 
non-trivial topologies.  One can obtain these black holes by either
constructing higher-dimensional counterparts of the identifications in
anti-de\thinspace Sitter spacetime \cite{man-amn,man-ban} or by systematic
elimination of the conical singularities in the cosmological C-metric
\cite{man-adss}.  

In this article I shall describe the basic properties of these topological black
holes from both an exterior and an interior viewpoint. I shall first describe 
their basic construction in terms of identifications on anti-de\thinspace Sitter 
spacetime.  I shall then move on to outline some of their basic properties,
and discuss how they can form from a distribution of collapsing dust. I shall
then explore the behaviour of the radiative falloff of waves outside these
black hole, a necessary precursor to obtaining a more detailed understanding
of their interior structure.  Finally, I shall close with a discussion of
the possible relevance of topological black holes to physics and of 
some recent research into this subject.

\section{Describing Topological Black Holes}

Topological black holes may be constructed by beginning with 
anti-de\thinspace Sitter spacetime and then judiciously imposing certain 
identifications which have the effect of generating event horizons 
in the resultant quotient spacetime, thereby yielding a black hole. The
treatment given here has not appeared before in the literature, although
various aspects of certain details have \cite{man-amn,man-ban,man-adss}.

Start with $(n+1)$ dimensional anti-de\thinspace Sitter spacetime. This can
be described by a hypersurface in an $(n+2)$-dimensional flat space of
signature $(n-2,2)$
\begin{equation}\label{man-1}
ds^2 = -dx^2_0 + dx^2_1 + \cdots + dx^2_{m-1} + dx^2_m + \cdots + dx^2_n - dx^2_{n+1}
\end{equation}
where the equation of constraint for the hypersurface is
\begin{equation}\label{man-2}
 -x^2_0 + x^2_1 + \cdots + x^2_{m-1} + x^2_m + \cdots + x^2_n - x^2_{n+1} = -\ell^2
\end{equation}
where $\ell$ is constant.

Consider next the subspace described by the coordinates $(x_0,\ldots,x_m)$. This
$m$-dimensional Minkowski subspace has the isometry group SO(m-1,1), with the
associated Killing vectors
\begin{equation}\label{man-3}
J_{\alpha\beta} = -J_{\beta\alpha}
= \left(x_\alpha \partial_\beta \pm x_\beta \partial_\alpha\right)
\end{equation}
where $\alpha,\beta = 0, \ldots, m-1$, and the plus sign is chosen if 
one of $\alpha$ or $\beta$ is $0$. The idea now is to identify points in this
subspace which are connected by some discrete subgroup $\Gamma$ of the 
isometries which are generated by the $J_{\alpha\beta}$. This of course can
be done in a variety of ways for the many different discrete subgroups.
However most of these identifications will yield closed timelike curves (CTCs) 
because points in the $x_0$ direction will end up being identified.

Hence the identification subspaces must lie in a region where the 
Killing vectors are spacelike to avoid CTCs.  This yields the constraint
\begin{equation}\label{man-4}
 x^2_0 - \left( x^2_1 + \cdots + x^2_{m-1}\right) = R^2 > 0
\end{equation}
on the Minkowskian subspace coordinates.  Reparametrizing the coordinates so
that
\begin{equation}\label{man-5}
 R X_{\alpha} = \frac{R_+}{\ell} x_{\alpha} 
\end{equation}
(where $R_+$ is an arbitrary constant)
implies that the $m$-dimensional Minkowski subspace metric may be written as
\begin{eqnarray}
ds^2_m &=& -dx^2_0 + dx^2_1 + \cdots + dx^2_{m-1} \nonumber\\
&=& \frac{\ell^2}{R^2_+}\left[-dR^2
+ R^2 \left(-dX^2_0 + dX^2_1 + \cdots + dX^2_{m-1}\right)\right]
\label{man-6}
\end{eqnarray}
where 
\begin{equation}\label{man-7}
 -X^2_0 + X^2_1 + \cdots + X^2_{m-1} = - \frac{R^2_+}{\ell^2} \qquad .
\end{equation}
This latter constraint means that we can write the quotient subspace metric as
\begin{equation}\label{man-8}
ds^2_m = -\frac{\ell^2}{R^2_+}dR^2 + R^2 d\sigma^2_{m-1}
\end{equation}
after identification.  The metric 
\begin{equation}\label{man-9}
d\sigma^2_{m-1} = \frac{\ell^2}{R^2_+}\left(-dX^2_0 + dX^2_1 + \cdots + dX^2_{m-1}\right)
\end{equation}
with the constraint (\ref{man-7}) describes, after identification, a compact 
$(m-1)$-dimensional space $\Sigma$ of negative curvature, with 
$\Sigma = H^{m-1}/\Gamma$.  

The coordinate $R$ is timelike within the $m$-dimensional subspace, but within the
full spacetime is actually timelike.  The full metric may now be written as
\begin{equation}\label{man-10}
ds^2 = -\frac{\ell^2}{R^2_+}dR^2 + R^2 d\sigma^2_{m-1}
+ dx^2_m + \cdots + dx^2_n - dx^2_{n+1}
\end{equation}
where the constraint (\ref{man-2}) becomes
\begin{equation}\label{man-11}
  x^2_m + \cdots + x^2_n - x^2_{n+1} = \ell^2\left(\frac{R^2}{R^2_+}-1\right) \qquad .
\end{equation}
From (\ref{man-11}) is clear that the above identification procedure has 
broken up the original anti-de\thinspace Sitter spacetime into various causal regions
that are parametrized by the magnitude of $R$.  When $R=R_+$ the coordinates
$(x_m,\ldots,x_{n+1})$ describe a null hypersurface.  This will later be seen to
be the event horizon of the black hole.  The region $R > R_+$ will correspond to
the exterior of the black hole, and the region $R < R_+$ will correspond to the
black hole interior, with $R=0$ being a singular point where the identification
procedure becomes degenerate, the condition (\ref{man-4}) being violated.

To see that the metric (\ref{man-10}) actually describes a black hole, it is helpful
to reparametrize the remaining coordinates
\begin{equation}\label{man-12}
\sqrt{R^2-R^2_+} Y_{I} = \frac{R_+}{\ell} x_{n+1-I}  \qquad \qquad I=m,\ldots,n+1
\end{equation}
in which case (\ref{man-11}) becomes
\begin{equation}\label{man-13}
 Y^2_1 + \cdots + Y^2_{n-m+1}-Y^2_0 = 1 
\end{equation}
in the region $R > R_+$. The orthogonal subspace metric may be written as
\begin{eqnarray}
ds^2_{n-m+1} &=& dx^2_m + \cdots + dx^2_{n+1} \\
&=& \frac{\ell^2}{R^2_+}\left[\frac{R^2}{R^2-R^2_+}dR^2
+ (R^2-R^2_+)\left(-dY^2_0 + dY^2_1 + \cdots + dY^2_{n-m+1}\right)\right]
\nonumber
\label{man-14}
\end{eqnarray}
in terms of the $Y$-coordinates and $R$. The metric in the spatial $Y$
coordinates may be rewritten in terms of  $(n-m+1)$ dimensional spherical coordinates
\begin{equation}\label{man-15}
dY^2_1 + \cdots + dY^2_{n-m+1} = d\rho^2 + \rho^2 d\Omega^2_{n-m}
\end{equation}
and the constraint (\ref{man-13}) becomes
\begin{equation}\label{man-16}
\rho^2 - Y^2_0 = 1  \qquad .
\end{equation}
This suggests the coordinate transformation
\begin{equation}\label{man-17}
\rho = \cosh\left(\frac{R_+}{\ell^2}t\right) \qquad 
Y_0 = \sinh\left(\frac{R_+}{\ell^2}t\right)
\end{equation}
which yields
\begin{equation}\label{man-18}
ds^2_{n-m+1} = \frac{\ell^2}{R^2_+}\left[\frac{R^2}{R^2-R^2_+}dR^2
+ (R^2-R^2_+)\left(- \frac{R^2_+}{\ell^4}dt^2 
+\cosh^2\left(\frac{R_+}{\ell^2}t\right)d\Omega^2_{n-m}\right)\right]
\end{equation}
for the orthogonal subspace metric.

Inserting (\ref{man-18}) into (\ref{man-10}) yields the final result
\begin{equation}\label{man-19}
ds^2 = \frac{\ell^4}{R^2_+} N(R) \left(- \frac{R^2_+}{\ell^4}dt^2 
+\cosh^2\left(\frac{R_+}{\ell^2}t\right)d\Omega^2_{n-m}\right)
+\frac{dR^2}{N(R)}  + R^2 d\sigma^2_{m-1}
\end{equation}
which is the general metric for an $(n+1)$-dimensional topological black hole.
The function $N(R)$ is given by 
\begin{equation}\label{man-20}
N(R) = \frac{R^2-R^2_+}{\ell^2}
\end{equation}
and plays the role of a generalized lapse function in the metric (\ref{man-19}).

The metric (\ref{man-19}) is an exact solution to the Einstein equations with
negative cosmological constant in $(n+1)$ dimensions. 
The event horizon at $R=R_+$ is the direct product of null $(n-m)$-sphere
with $\Sigma$, a generalization of the usual event horizon of
a spherically symmetric black hole in $(3+1)$ dimensions which is a direct
product of a pair of crossed null lines ({\it i.e.} a null $0$-sphere) with
a $2$-sphere.  Similarly, using (\ref{man-4}), (\ref{man-10}), 
and (\ref{man-11}), one can see
that there is a singularity at $R=0$ which is the direct product of a 
hyperbolic surface with $\Sigma$, generalizing the usual singularity which
is the direct product of a hyperbola with a $2$-sphere.  

It is easier to appreciate some of the features of this metric by considering a
few special cases.  
\begin{itemize}
\item{$m=n=2 \quad$} The metric is
\begin{equation}\label{man-21}
ds^2 = -N(R) dt^2 + \frac{dR^2}{N(R)}  + R^2 d\phi^2 
\end{equation}
which is the static BTZ metric.  The compact space $\Sigma$ is a circle.
The event horizon is the direct product of a pair of crossed null lines with
this circle, and the singularity at $R=0$ is the direct product of
a hyperbola with this circle.

\item{$m=n=3 \quad$} The metric is now
\begin{equation}\label{man-22}
ds^2 = -N(R) dt^2 + \frac{dR^2}{N(R)} 
+ R^2\left(d\theta^2 + \sinh^2(\theta) d\phi^2\right)
\end{equation}
which are the $(3+1)$ dimensional topological black holes
discussed in refs. \cite{man-amn,man-adss}.
The space $\Sigma = \Sigma_g$ is a Riemann 2-surface of genus $g$.
The event horizon is the direct product of a pair of crossed null lines with
$\Sigma_g$ and the singularity at $R=0$ is the direct product of
a hyperbola with $\Sigma_g$.

The compactness of $\Sigma_g$ can be understood in the following way.
The coordinates $(\theta,\phi)$ are the coordinates of a hyperbolic space or
pseudosphere. Geodesics on the pseudosphere are formed  from intersections of the
psuedosphere with planes through the origin, and are the analogs of
great circles on  a surface of
constant positive curvature (a sphere), which are intersections of
the sphere and planes through the origin.
A projection of the psuedosphere onto the $(y_1, y_2)$ plane
is known as a the Poincar\'e disk. On it, geodesics
are segments of circles, orthogonal to the disk boundary at the edges.  
The pseudosphere, its associated
Poincar\'e disk and the geodesics are shown in figure \ref{man-F1}. 
  \begin{figure}
    \leavevmode
    \hfil \hbox{\epsfysize=7cm \epsffile{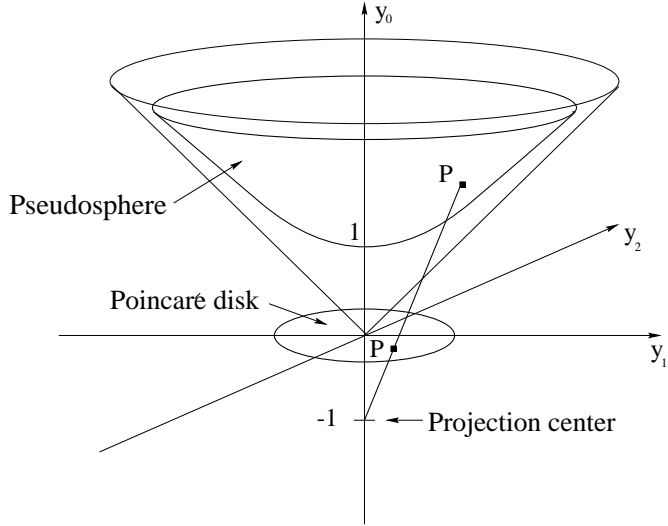}} \hfil
    \caption{The pseudosphere is one half of the hyperboloid. 
Beneath the pseudosphere is the Poincar\'e disk, the 
center of which is the origin.}
    \label{man-F1}
  \end{figure}
Consider a polygon centered at the origin of the pseudosphere whose sides are geodesics.
By identifying opposite sides of this polygon a compact surface on the pseudosphere 
can be obtained. The angles of the polygon must sum to
$2 \pi$ or more, and the number of sides must be a multiple of four in order to 
avoid conical singularities \cite{man-Bal}.
Since the geodesics on the pseudosphere meet at angles smaller than those for geodesics
meeting on a flat plane, an octagon is the simplest solution, yielding a surface of
genus 2.  This construction is shown in figure \ref{man-F2}.  In general, a polygon of $4g$ sides
yields a surface of genus $g$, where $g\geq 2$.
  \begin{figure}
    \leavevmode
    \hfil \hbox{\epsfysize=10cm \epsffile{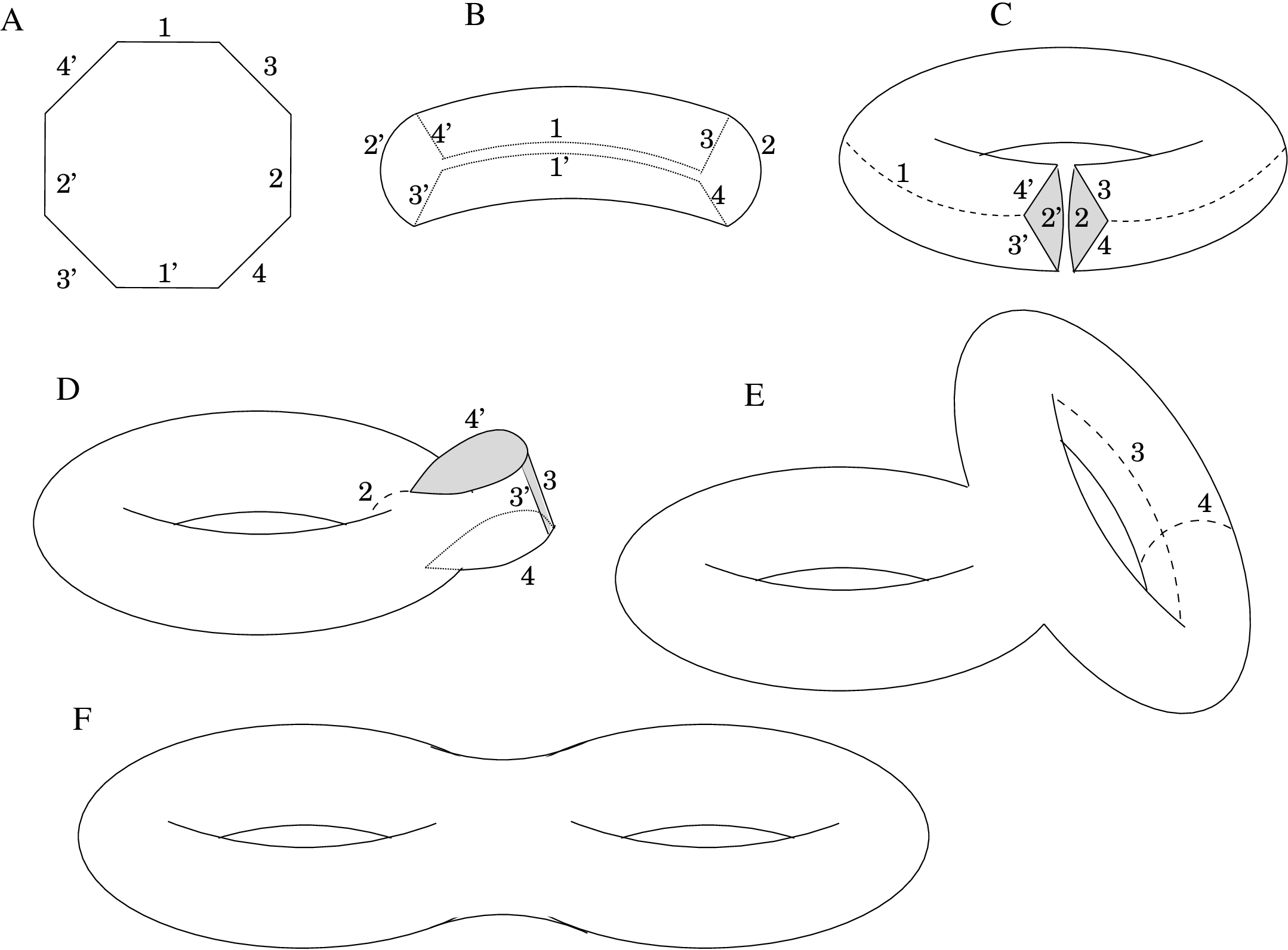}} \hfil
    \caption{The identification of the octagon is shown. Opposite sides of the octagon 
are identified as in 2A, with sides drawn straight for clarity.  Dashed lines indicate 
where sides have been sewn together.   Sides $1$ and $ 1^\prime$ are first identified, 
folding the top and  bottom of the octagon away from view (2B).  The sides $2$ and 
$2^\prime$ are then brought together to  form a torus with a
diamond shaped  hole, as in 2C.  Next, sides $3$ and $3^\prime$ are stretched out and joined 
in 2D.  The loop is lengthened along the direction of identification 3 and bent until $4$ and 
$4^\prime$ meet, forming a second torus.  Finally the topology is deformed to the preferred 
shape, seen in 2F.  Identification of a polygon of $4g$ sides will result
in $g$ attached tori or, equivalently, a $g$-holed pacifier.}
\label{man-F2}
 \end{figure}

\item{$m=2,n=3 \quad$} The metric is
\begin{equation}\label{man-23}
ds^2 = N(R) \left(-dt^2 + \frac{\ell^4}{R^2_+} 
\cosh^2\left(\frac{R_+}{\ell^2}t\right)d\theta^2\right)
+\frac{dR^2}{N(R)}  + R^2 d\phi^2
\end{equation}
which describes the metric of an alternate generalization of the BTZ black
hole discussed recently by Holst and Peldan \cite{man-hol}. The coordinate
$\theta$ is periodically identified. The compact
space $\Sigma$ is again a circle with coordinate $\phi$. The event horizon
is the direct product of a null conoid ({\it i.e.} a null circle)
with this circle, and the singularity is the direct product of a 2-dimensional
hyperboloid with this circle.  

Under the coordinate transformation
\begin{equation}\label{man-24}
\sinh(\frac{R_+ t}{\ell^2}) = \cosh(T/\ell) \sqrt{X^2-1}\qquad   
\tan(\theta) = \sinh(T/\ell) \frac{\sqrt{X^2-1}}{X}
\end{equation}
the $(t,\theta)$ section transforms as
\begin{equation}\label{man-25}
-dt^2 + \frac{\ell^4}{R^2_+} 
\cosh^2\left(\frac{R_+}{\ell^2}t\right)d\theta^2 =
\frac{\ell^4}{R^2_+}\left(-\frac{dX^2}{X^2-1}+ (X^2-1)dT^2/\ell^2\right)
\end{equation}
where the coordinate transformation is valid provided $|X| > 1$.  It is
clear from (\ref{man-25}) that these coordinates can be extended to the $|X|\leq 1$ region,
in which $X$ becomes a spatial coordinate and $T$ becomes a timelike coordinate. Writing
$X = \cos\lambda$ yields
\begin{equation}\label{man-26}
\frac{\ell^4}{R^2_+}\left(-\frac{dX^2}{X^2-1}+ (X^2-1)dT^2/\ell^2\right)
= \frac{\ell^4}{R^2_+}\left(d\lambda^2 - \sin^2\lambda\ dT^2/\ell^2\right)
\end{equation}
which implies that
\begin{equation}\label{man-27}
ds^2 = N(R) 
\frac{\ell^4}{R^2_+}\left(d\lambda^2 - \sin^2\lambda\ dT^2/\ell^2\right)
+\frac{dR^2}{N(R)}  + R^2 d\phi^2
\end{equation}
which is also a solution to the Einstein equations with negative cosmological constant.

The metric (\ref{man-27}) is the constant curvature black hole (CCBH) discussed
earlier \cite{man-ban}.  Its global properties differ from those of the metric
(\ref{man-23}) in that the event horizon
is the direct product of a pair of null conoids joined at their apexes
with the circle $\Sigma$.  Note that $\partial/\partial T$ is a Killing vector of
the metric (\ref{man-27}), but that $\partial/\partial t$ is not a Killing vector of
the metric (\ref{man-23}).  In this sense the event horizon of (\ref{man-23}) evolves with 
respect to the time coordinate $t$ \cite{man-hol}.

\end{itemize}

\section{Properties of Topological Black Holes}

I shall consider only $(3+1)$ dimensional black holes throughout the sequel.  For the
solutions (\ref{man-22}) it is possible to add mass and charge \cite{man-adss}, yielding
\begin{eqnarray}
ds^2 &=& -\left(R^2/l^2-1-2m/R+q^2/R^2\right)dt^2 + \frac{dR^2}{R^2/l^2-1-2m/R+q^2/R^2}
   \nonumber\\ 
 &&\qquad + R^2\left(d\theta^2 + \sinh^2(\theta) d\phi^2\right)\label{man-28}
\end{eqnarray}
which is an exact solution of the Einstein-Maxwell equations with negative cosmological
constant $\Lambda=-3/\ell^2$.  The metric for fixed $(t,R)$ is assumed to be identified in
the $(\theta,\phi)$ coordinates as decribed previously, so that it describes a black hole
whose event horizon is of genus $g\geq 2$.  The electromagnetic field strength is
\begin{equation}\label{man-29}
F = -\frac{q}{R^2} dt \wedge dR
\end{equation}
in the electric case, and 
\begin{equation}\label{man-30}
F = q\sinh\theta d\theta \wedge d\phi 
\end{equation}
in the magnetic case.  Toroidal black holes (genus $g=1$) have the metric
\begin{eqnarray}
ds^2 = -\left(R^2/l^2-2m/R+q^2/R^2\right)dt^2 + \frac{dR^2}{R^2/l^2-2m/R+q^2/R^2} 
\nonumber \\
+ R^2\left(d\theta^2 + d\phi^2\right) \label{man-31}
\end{eqnarray}
with 
\begin{equation}\label{man-32}
F = -\frac{q}{R^2} dt \wedge dR  \qquad F = q d\theta \wedge d\phi 
\end{equation}
in the electric and magnetic cases respectively, where $\theta$ and $\phi$ are
periodically identified.  Note that the entire spacetime has 
topology $R^2\times H^2_g$ for a genus $g\geq 1$ black hole.

Using the quasilocal formalism developed for anti de Sitter spacetimes \cite{man-bjm}
it is straightforward to show that 
\begin{equation}\label{man-33}
M =   m(|g-1|+\delta_{g,1})
\end{equation}
is the conserved mass parameter associated with the Killing vector $\partial/\partial t$.
for genus $g\geq 1$ \cite{man-adsl,man-BLP}. Similarly, 
\begin{equation}\label{man-34}
Q = q(|g-1|+\delta_{g,1})
\end{equation}
is the conserved charge $Q$ associated with a genus $g$ black hole.

The genus $g$ metric function 
\begin{equation}\label{man-35}
V(r) \equiv R^2/l^2-(1-\delta_{g,1}-2\delta_{g,0})-2m/R+q^2/R^2
\end{equation}
has at most two roots for positive $R$, corresponding to an inner
and outer horizon, as with the usual $g=0$ 
Reissner-Nordstrom anti de\thinspace Sitter metric. For $g=1$,
provided 
\begin{equation}\label{man-36}
27\,l^{2}\,{\it m}^{4} \geq 16\,q^{6}
\end{equation}
there are two horizons, with the extremal case saturating the inequality. For 
$g\geq 2$ event horizons exist provided
\begin{equation}\label{man-37}
m^2 \leq 
  {\displaystyle \frac {l^2}{27}} \,{\displaystyle \frac { 16 -
24\,e^{2}b -16b\sqrt{1 - e^{2}b}\,e^{2} + 6b^2\,e^{4} 
+ 16\,\sqrt{1 - e^{2}b}}{e^{6}}} 
\end{equation}
where $e = \frac{2\sqrt{2}q}{3m}$. There is no
(obvious) upper limit on e, and event horizons can exist for arbitrarily
large values of $q$ relative to $m$.   

\begin{figure}[htbp]
    \hfill \psfig{file=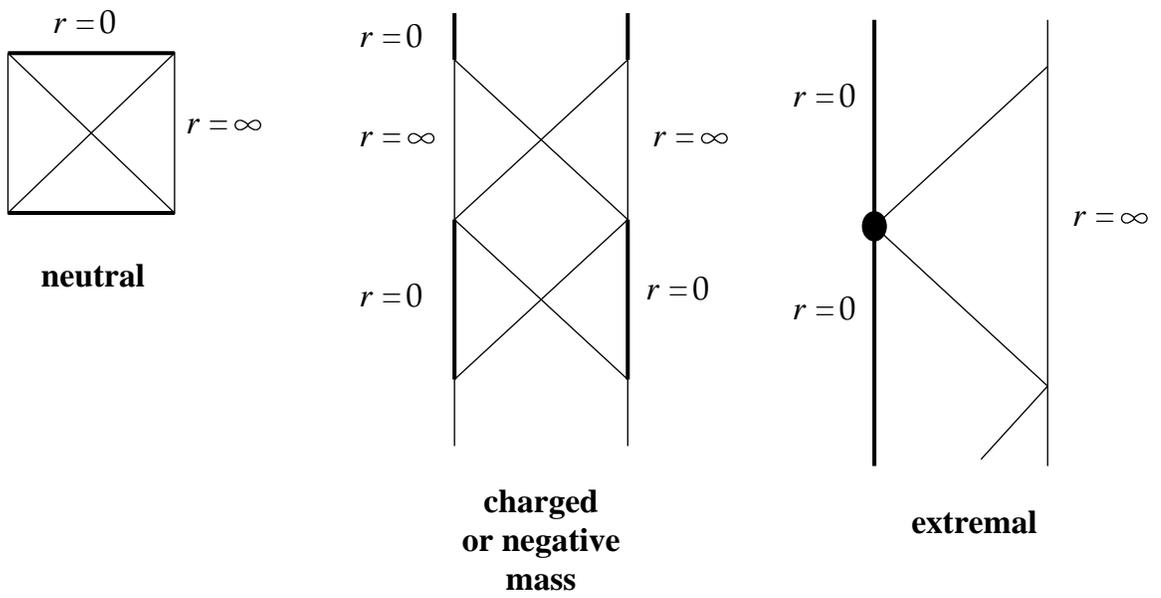,width=6in} \hfill \mbox{}
\caption{Causal diagrams for topological black holes in the neutral, subextremal
and extremal cases.  The subextremal case can be that of a charged black hole or
of a negative mass black hole.}
    \label{man-F3}
  \end{figure}

The causal structure of these spacetimes is similar to that of 
Reissner-Nordstrom anti de\thinspace Sitter spacetime with spherical topology.
The three causal diagrams in the neutral, sub-extremal and extremal cases
are shown in Fig. \ref{man-F3}.  A curious feature of these higher-genus black holes is that
the mass parameter need not be positive in order for an event horizon to exist
\cite{man-neg}, even if the black hole is uncharged.

An extensive study of the thermodynamic properties of these black holes was
recently carried out by Brill {\it et. al.} \cite{man-BLP}.  The entropy/area
relation
\begin{equation}\label{man-38}
S= \frac{1}{4} A
\end{equation}
was found to hold, although this result has been disputed by Vanzo \cite{man-vanzo}.
Their heat capacities $C_X = T \left(\frac{\partial S}{\partial T}\right)_X$ were also 
computed, and found to be of indefinite sign if $g\geq 2$.  

It has also been demonstrated that topological black holes of
arbitrary genus can form from the gravitational collapse of pressureless dust
\cite{man-wen}.  Although the procedure is somewhat analogous to that in the 
spherical case, there are a few interesting discrepancies. For non-trivial topologies,
collapse from rest will not take place unless
\begin{equation}\label{man-39}
\Lambda > 8 \pi G \rho_0,
\end{equation}
where $\rho_0$ is the initial density of the cloud.  A more general but
qualitatively similar condition holds if  the dust cloud is given some initial
velocity.  The exterior spacetime formed is that given by (\ref{man-28}) with
$q=0$.  Another interesting feature is that collapse can take place even
for dust which violates the weak energy condition, {\it i.e.} $\rho_0 < 0$.
Provided the initial negative energy density is not too large in magnitude,
a cloud of negative energy dust can collapse to a topological black hole of
negative mass \cite{man-neg}. 

Finally, topological black holes may be pair produced in the presence of a domain
wall of suitable topology \cite{man-adss,man-adsl}.  The mechanism is analogous 
to that discussed for other cosmological black holes \cite{man-Ros,man-gib}.

\section{Inside Topological Black Holes?}

Both charged and negative mass topological black holes have inner as well as outer
event horizons.  As can be seen from figure \ref{man-F3}, 
the maximal extensions of such
black hole spacetimes can be imagined as a collection of different
asymptotically anti de\thinspace Sitter universes connected by different charged
(or negative mass)  black holes. The presence of the inner horizon causes
any radiation (either scalar, electromagnetic or gravitational in nature) 
entering this kind of black hole to be  indefinitely blue-shifted at the inner 
(or Cauchy) horizon. 

Does the phenomenon of mass inflation \cite{man-poi} take place for topological
black holes?  There are in general three necessary ingredients for mass inflation
to take place: the existence of a Cauchy horizon, the presence of radiation
which enters the black hole at some rate which decays at late times, and the
possibility of cross-flow of this radiation with that emitted
from a collapsing object that is forming the black hole.  The first 
condition is clearly satisfied by either charged or negative mass black holes.
Since both kinds of black holes can be formed from gravitational collapse,
it is clear that the third condition can be satisfied provided the second one is.

An understanding of the nature of radiative falloff outside topological black holes 
is therefore crucial insofar as investigating their interior is concerned.
For asymptotically flat spacetimes, the radiative falloff at late times obeys
a power-law \cite{man-Pri}. However this situation changes for other kinds of
spacetimes. For example Mellor and Moss have shown that the radiation from 
perturbations in a de\thinspace Sitter background exponentially 
decreases \cite{man-Mel}. Strictly
  speaking this result has nothing to do with late time falloff
  since the global geometry extends beyond the cosmological
  horizon. However it does indicate that the radiative falloff
  behaviour is sensitive to the asymptotic structure of the spacetime.
Indeed, for conditions more general that simple asymptotic flatness,
the tail can be something other than an inverse power-law \cite{man-chi1}.

A recent investigation \cite{man-fal} of radiative falloff in 
Schwarzchild anti de\thinspace Sitter spacetime indicated  that the 
late time falloff behaviour is considerably more complicated than that in 
either asymptotically flat or asymptotically de\thinspace Sitter spacetimes.  
This is a consequence of spatial infinity being timelike -- although the
proper distance to any point at large $R$ in such a spacetime diverges as
$R\to \infty$, a light ray can travel to arbitrarily large $R$ in a finite
amount of time.  

A useful means for probing the falloff behaviour of late times is to
study the  conformally coupled scalar wave equation.  The scalar wave
equation has qualitatively the same behaviour and its more complicated
electromagnetic and gravitational tensorial counterparts, and the conformal
coupling ensures that the effective potential in which the wave moves
most closely resembles the previously studied asymptotically flat case
\cite{man-fal}.  

The (conformally coupled) scalar wave equation in $(3+1)$ dimensions is
  \begin{eqnarray}
    \Del^2 \Psi & = & \xi\,R\,\Psi \comma \label{man-40}
  \end{eqnarray}
  where $\xi$ is an arbitrary constant. If $\xi = \frac{1}{6}$
  this equation is conformally invariant.  The form of the metric
  for topological black holes is 
  \begin{eqnarray}
    ds^2 & = &
    \minus V(r)\,dt^2 + \frac{dr^2}{V(r)} + r^2\,d{\Omega_g}^2
    \comma \label{man-41}
  \end{eqnarray}
  where $V(r)$ is the lapse function given by (\ref{man-35})
  and $d{\Omega_{g}}^2$ is the metric of a genus $g$ Riemann surface.
  Assuming the separability condition 
  \begin{eqnarray}\label{man-42}
    \Psi & = & \frac{1}{r}\,\psi(t,r)\,{\cal Y}(\Omega_g)_l 
  \end{eqnarray}
it is straightforward to show that equation 
(\ref{man-40}) gives
  \begin{eqnarray}
    \minus \di_{tt} \psi(t,r) + V(r)\,\dir \SqBk{N(r)\,\dir \psi(t,r)}
    - V(r)\,V_e(r)\,\psi(t,r)
    & = &  0  \label{man-43}
  \end{eqnarray}
  where
  \begin{eqnarray}
    V_e(r) & \equiv &
    \xi\,R + \frac{1}{r}\,\frac{d}{d\,r} V(r)
    + \frac{l(l+1)}{r^2}  \label{man-44}
  \end{eqnarray}
defines the function $V_e(r)$.  The functions ${\cal Y}_l$ are the 
genus $g$ analogues of the spherical harmonics \cite{man-Bal}, and
for $g\geq 2$, say, satisfy the equation
  \begin{equation}\label{man-45}
    \hat{L}^2\!\SqBk{{\cal Y}_l} \equiv
\left[\frac{1}{\sinh\theta}\frac{\partial}{\partial\theta}
\left(\sinh\theta\frac{\partial}{\partial\theta}\right) + 
\frac{1}{\sinh^2\theta}\frac{\partial^2}{\partial\phi^2}\right]{\cal Y}_l
=  \minus l\,(l+1)\,{\cal Y}_l 
\end{equation}
and are referred to as conical functions. 

For simplicity the scalar wave will
be assumed to be independent of $(\theta,\phi)$.
One can then rewrite the wave equation (\ref{man-43}) as
  \begin{eqnarray}
    \di_{tt} \psi\!\RdBk{t,r(x)} - \di_{xx} \psi\!\RdBk{t,r(x)}
    + {\cal V}\!\RdBk{r(x)}\,\psi\!\RdBk{t,r(x)} & = & 0 \label{man-46}
  \end{eqnarray}
where  ${\cal V}(x) \equiv V(r(x))V_e(R(x))$ with $l=0$ and 
  \begin{equation}\label{man-47}
    x \; \equiv \; \int \frac{dr}{N(r)} 
  \end{equation}
is the so-called tortoise coordinate.

The function  ${\cal V}(r)$ plays the role of
  a potential barrier which is induced from the background
  spacetime geometry. Equation (\ref{man-46}) has the familiar form of a potential
  scattering problem, although ${\cal V}$ has a rather complicated dependence
on the tortoise co-ordinate $x$. It can be integrated numerically in a
  straightforward fashion by using finite difference methods. The D'Alembert 
  operator $\di_{tt} - \di_{xx}$ is first discretized as
  \begin{eqnarray}\label{man-48}
    \frac{\psi(t-\Delta t,x)-2\,\psi(t,x)+\psi(t+\Delta t,x)}{{\Delta t}^2}\qquad&&\\
    \qquad - \frac{\psi(t,x-\Delta x)-2\,\psi(t,x)+\psi(t,x+\Delta x)}{{\Delta x}^2}
    &+& O({\Delta t}^2) + O({\Delta x}^2)\nonumber
  \end{eqnarray}
  using Taylor's theorem. In order to formulate a well-posed
  Cauchy problem initial conditions must be chosen. For simplicity these
can be taken to be
\begin{eqnarray}\label{man-49}
    \psi(t=0,x) \; = \; 0
    \hspace{1cm} & {\rm and} & \hspace{1cm}
    \di_t \psi(t=0,x) \; = \; u(x) \period
  \end{eqnarray}
  Because the field $\psi$ is initially zero, its subsequent
  evolution is solely the result of the initial impulse of the
  field $\di_t \psi$. Discretizing the second condition in
  (\ref{man-49}) yields
  \begin{eqnarray}\label{man-50}
    \frac{\psi(\Delta t,x)-\psi(\minus \Delta t,x)}{2\,\Delta t}
    & = & u(x) + O({\Delta t}^2) \comma
  \end{eqnarray}
  where a Gaussian distribution with finite support for
  $u(x)$ is employed. Defining
  \begin{eqnarray}
    \psi(m\,\Delta t,n\,\Delta x) & \equiv & \psi_{m,n} \comma \label{man-51}\\
    V(n\,\Delta x) & \equiv & V_n \comma \label{man-52}\\
    u(n\,\Delta x) & \equiv & u_n  \comma \label{man-53}
  \end{eqnarray}
  where the mesh size has to satisfy the condition
  $\Delta x > \Delta t$ so that the numerical rate of propagation
  of data is greater than its analytical counterpart. 

  If the black hole geometry is asymptotically
  flat, the tortoise coordinate $x$ goes from negative infinity to
  positive infinity. When the background is asymptotically
  anti de\thinspace Sitter the initial data no longer
  enjoy this privilege because the tortoise coordinate goes from minus
  infinity to zero only. In other words, rightward propagating
  data cannot travel in this direction forever. As with the semi-infinite 
  vibrating string problem, boundary conditions at
  spatial infinity (i.e. $x = 0$) are needed in the asymptotically
  anti de\thinspace Sitter background in order to formulate the problem
  appropriately. Here the boundary conditions employed will be
either
  \begin{eqnarray}
    \psi(t,x=0) \; = \; 0
    \hspace{1cm} & {\rm and} & \hspace{1cm}
    \di_x \psi(t,x=0) \; = \; 1 \label{man-54}
  \end{eqnarray}
which are referred to as the Dirichlet boundary conditions, 
or
  \begin{eqnarray}
    \psi(t,x=0) \; = \; 1
    \hspace{1cm} & {\rm and} & \hspace{1cm}
    \di_x \psi(t,x=0) \; = \; 0 \period \label{man-55}
  \end{eqnarray}
which are the Neumann boundary conditions.

Both of these boundary conditions will be used in investigated the
radiative falloff of conformally coupled waves in the topological
black hole spacetimes (\ref{man-41}).  For simplicity, only 
neutral topological black hole spacetimes will be considered.

\section{Radiative Falloff Outside Neutral Topological Black Holes}

Before presenting the results of the numerical analysis of equation
\ref{man-46} it will be worthwhile recapitulating what takes place in
Schwarzchild and Schwarzchild anti de\thinspace Sitter spacetimes
\cite{man-fal}.  The wave equation (\ref{man-46}) in all cases is
solved numerically using the scheme discussed in the previous section,
and $\xi = \frac{1}{6}$ throughout.

\begin{figure}[htbp]
    \hfill \psfig{file=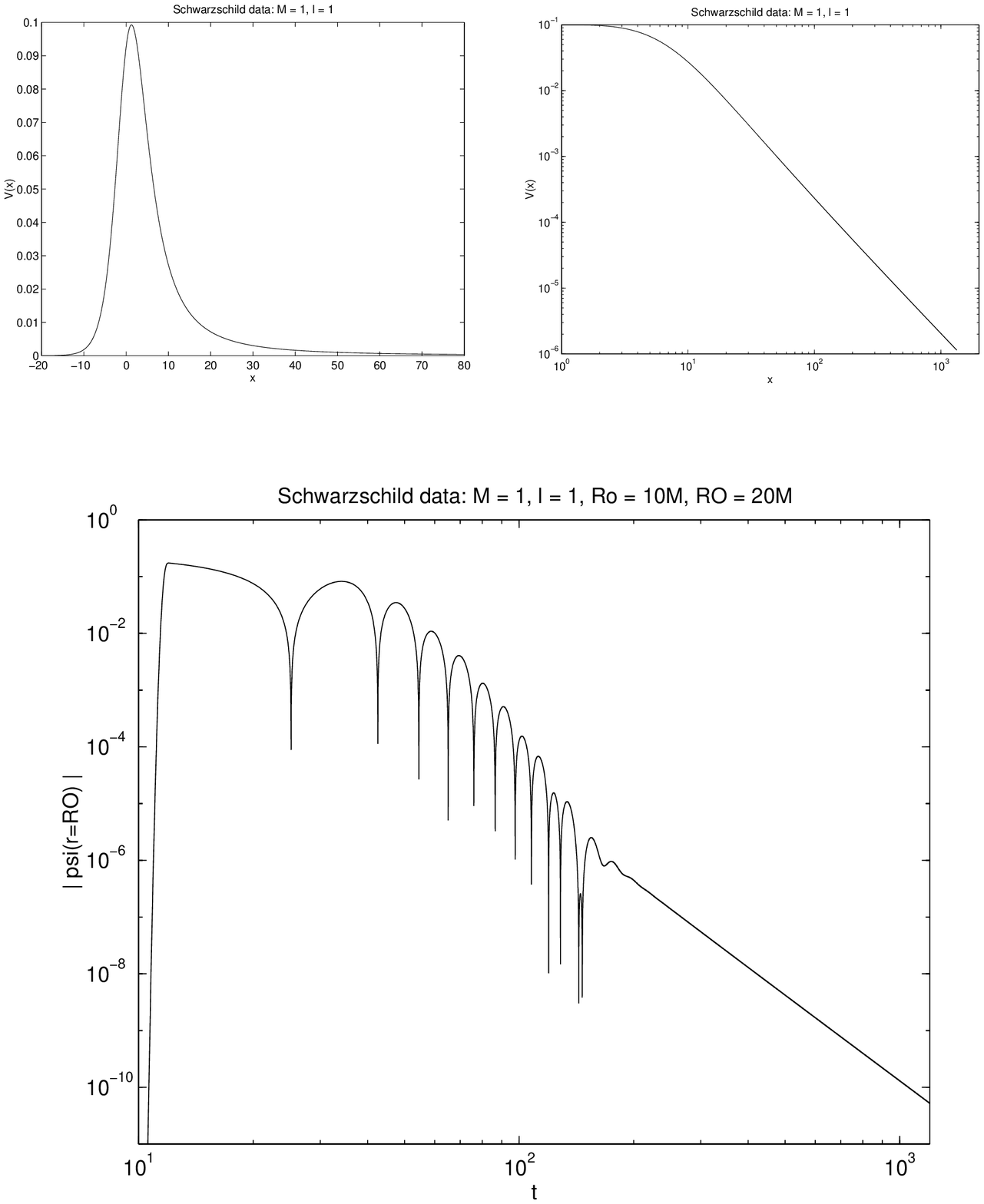,height=20cm} \hfill \mbox{}
    \caption{The potential ${\cal V}(x)$ in a Schwarzschild background
         and the resultant decay of a scalar wave.
	     Prior to $t\approx 200$ the decay is accompanied by `ringing' of
	     the quasi-normal modes, after which the falloff rate is
	     that of an inverse power-law.}
    \label{man-F4}
  \end{figure}

Figure \ref{man-F4} shows the general form of the potential 
${\cal V}(x)$ for a background Schwarzschild spacetime in both linear
and logarithmic coordinates (with the $l=1$ spherical harmonic).  
The bottom diagram in \ref{man-F4} is a logarithmic plot of the magnitude of the
scalar wave at the point $RO = 20 M$ as a function of time, where
the compact initial Gaussian pulse is at the point $Ro = 10 M$ (or $x = 12.76\, M$). 
The field vanishes 
until the pulse has propagated outwards to the point $RO$. It reaches
its maximum value, after which it undergoes a ``ringing'' effect due to the
presence of quasinormal modes \cite{man-chi1}.  This ringing dies out after
$t\approx 200$, after which the field decays according to a smooth power law
which from linear regression, is found to have a slope of $\minus 5.026$ in
  agreement with the analytic prediction of an inverse power-law
  falloff with exponent $2\,l+3$ \cite{man-Pri}. 

\begin{figure}[htbp]
    \hfill \psfig{file=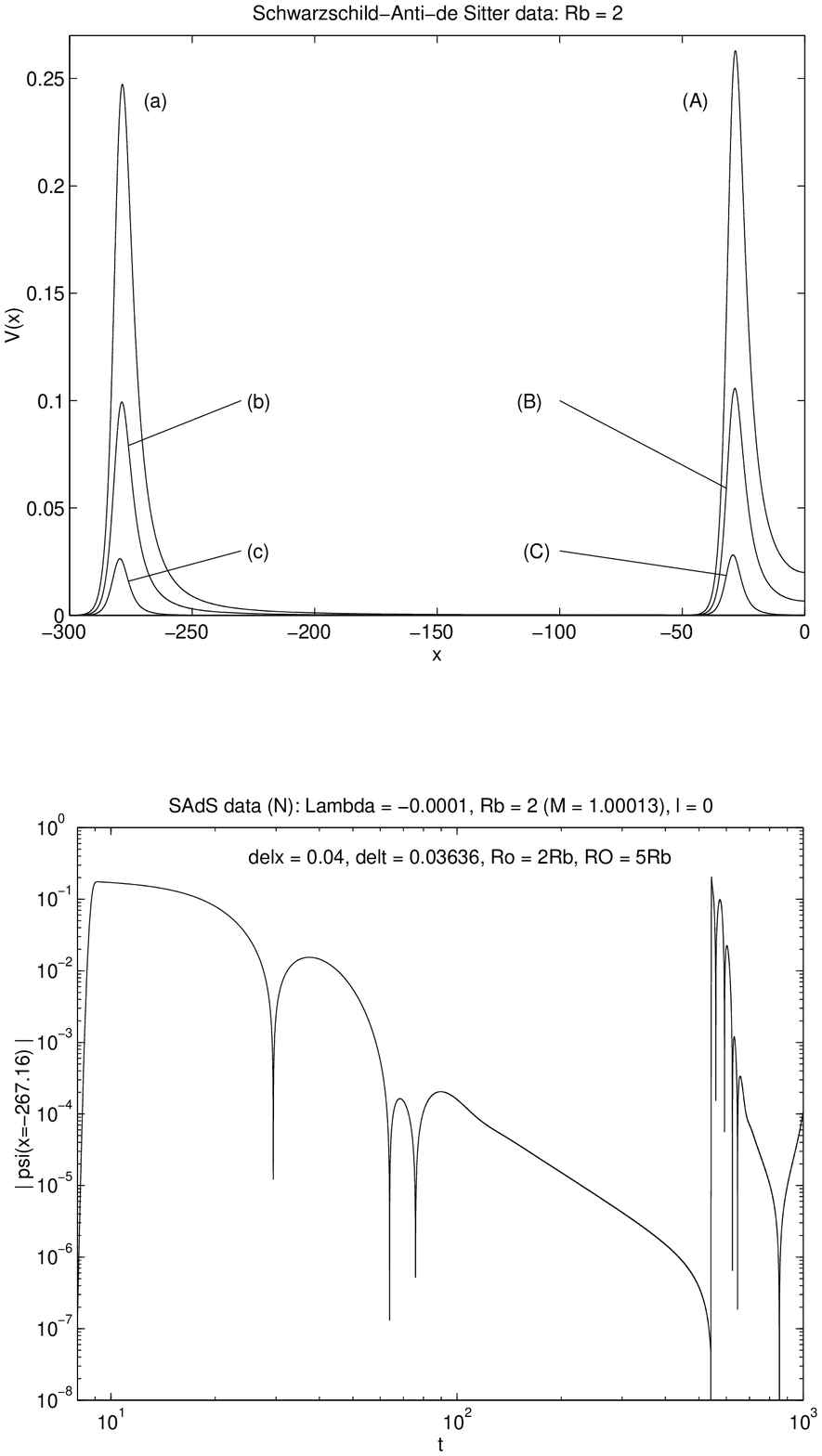,height=20cm} \hfill \mbox{}
    \caption{Potential functions $V(x)$ for the SAdS background.
	     The six potentials are generated with the parameters
	     (a) $\Lambda = \minus 10^{\minus 4}, l = 2$,
	     (b) $\Lambda = \minus 10^{\minus 4}, l = 1$,
	     (c) $\Lambda = \minus 10^{\minus 4}, l = 0$,
	     (A) $\Lambda = \minus 10^{\minus 2}, l = 2$,
	     (B) $\Lambda = \minus 10^{\minus 2}, l = 1$,
	     (C) $\Lambda = \minus 10^{\minus 2}, l = 0$. 
The bottom diagram shows the $l=0$ scalar wave falloff pattern using
		 Neumann condition.}
    \label{man-F5}
  \end{figure}

The situation for Schwarzchild anti de\thinspace Sitter (SAdS) spacetime is 
somewhat different, and is shown in figure \ref{man-F5}.  
As with the Schwarzschild black hole (figure \ref{man-F4}),
  the potential function $V(x)$ attains a maximum not far away
  from the event horizon $Rb$, given by the largest positive solution to
$V(R_b)=0$.   However unlike the Schwarzschild case, the tortoise coordinate $x$
  for the SAdS background is bounded above. The top part of
figure \ref{man-F5} illustrates the  shape of the
  potential function ${\cal V}(x)$ for a variety of values of 
$|\Lambda| = 3/\ell^2$ and the spherical harmonic parameter $l$.
Eventually all  the outgoing conformal waves that leave the black hole region will
  return towards it due to the boundary condition at $x = 0$.
  The returning wave will then reflect off of the potential
  barrier back toward spatial infinity for both the Dirichlet and Neumann
boundary conditions. In this and all subsequent diagrams
the quantities ``delx''  and ``delt'' on the graphs refer to the step sizes 
$\Delta x$ and $\Delta t$ of the variables x and t respectively, where 
$\Delta x > \Delta t$ holds as noted above.

\begin{figure}[htbp]
    \hfill \psfig{file=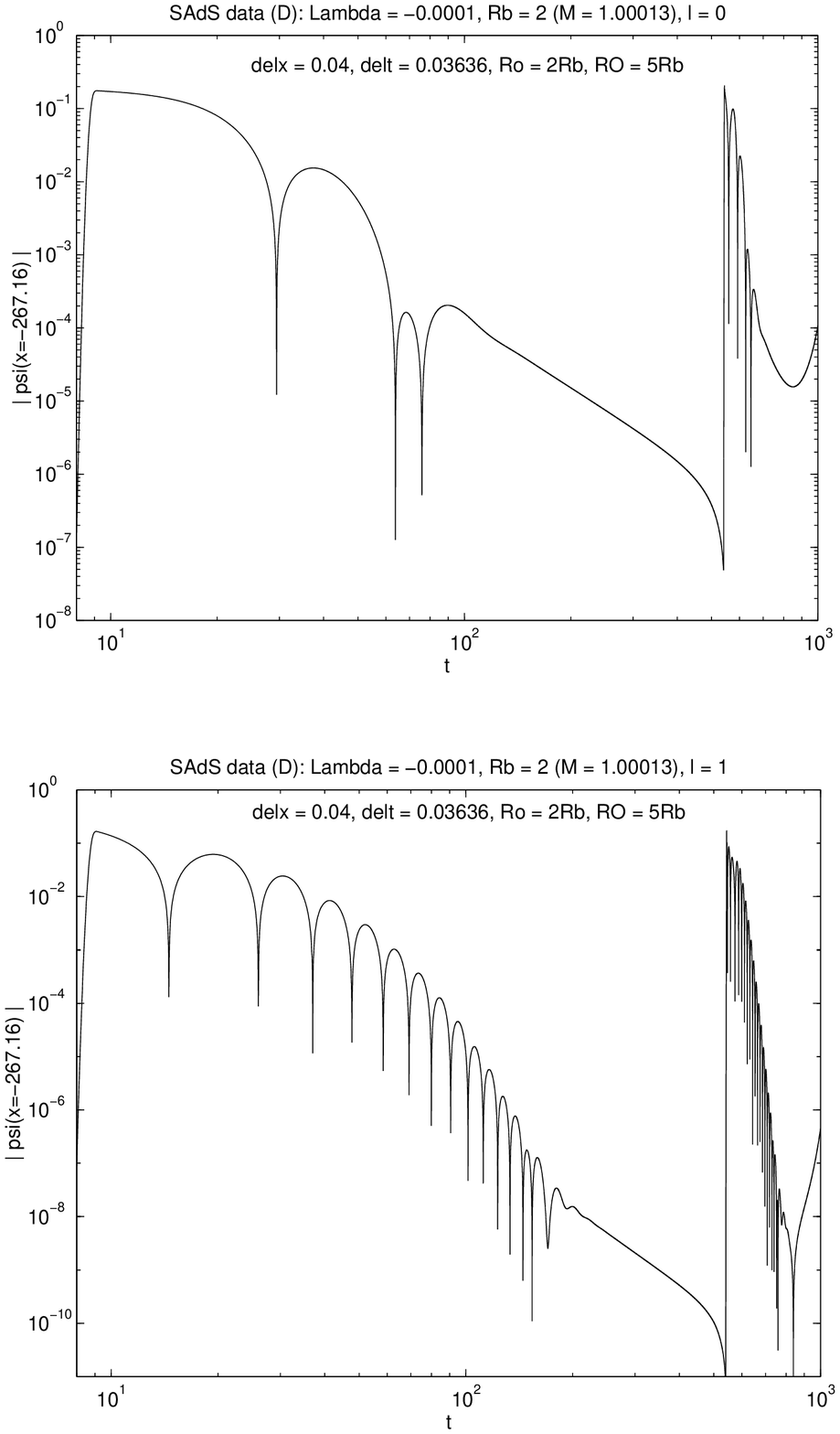,height=20cm} \hfill \mbox{}
    \caption{The $l=0$ (top) and $l=1$ (bottom) scalar wave falloff patterns using
		 the Dirichlet condition.}
    \label{man-F6}
  \end{figure}

\begin{figure}[htbp]
    \hfill \psfig{file=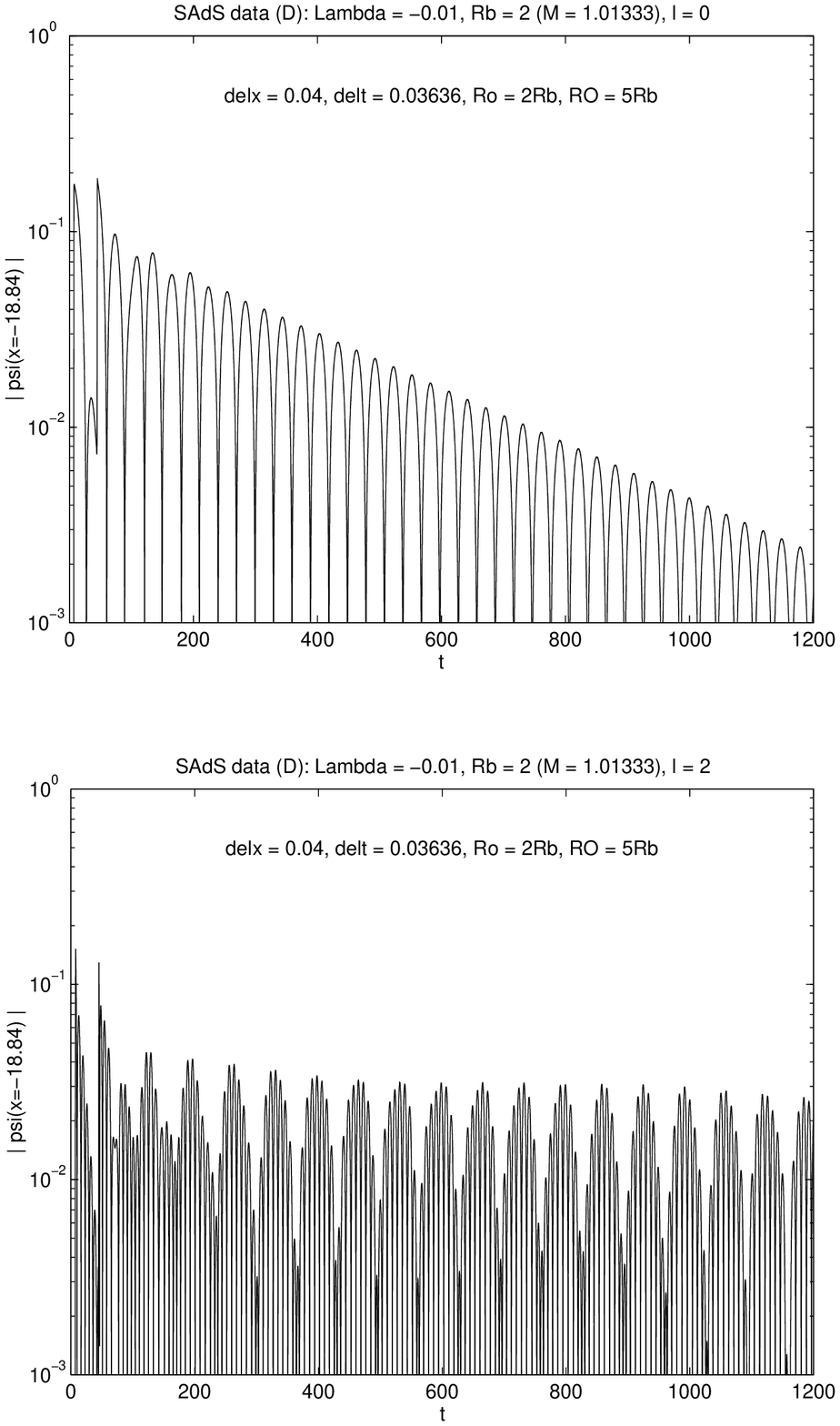,height=15cm} \hfill \mbox{}
    \caption{The $l=0$ (top) and $l=2$ (bottom) scalar wave falloff patterns using
		 for large $|\Lambda|$.}
    \label{man-F7}
  \end{figure}

For small $|\Lambda|$ and small times, the falloff behaviour
  resembles the Sch\-warz\-child case. Initially there is a ringing
  effect (due to the quasi-normal modes) followed by inverse
  power-decay behaviour. However this inverse power-decay does not
  last very long because of the return of the outgoing wave from
  spatial infinity.  As $l$ gets larger the ringing increases in
frequency as shown in figure \ref{man-F6}.
As $|\Lambda|$ increases, the transient power-law effect is eliminated entirely
and new behaviour emerges. For small $l$ there can be a pure ringing effect within
an exponentially decaying envelope, as shown in the top part of figure \ref{man-F7}.
As $l$ increases, the ringing itself undergoes a complicated oscillation effect
which mildly decays, as shown in the bottom part of figure \ref{man-F7}.
These results are all largely insensitive to the use of either Dirichlet or
Neumann boundary conditions \cite{man-fal}.

\begin{figure}[htbp]
    \hfill \psfig{file=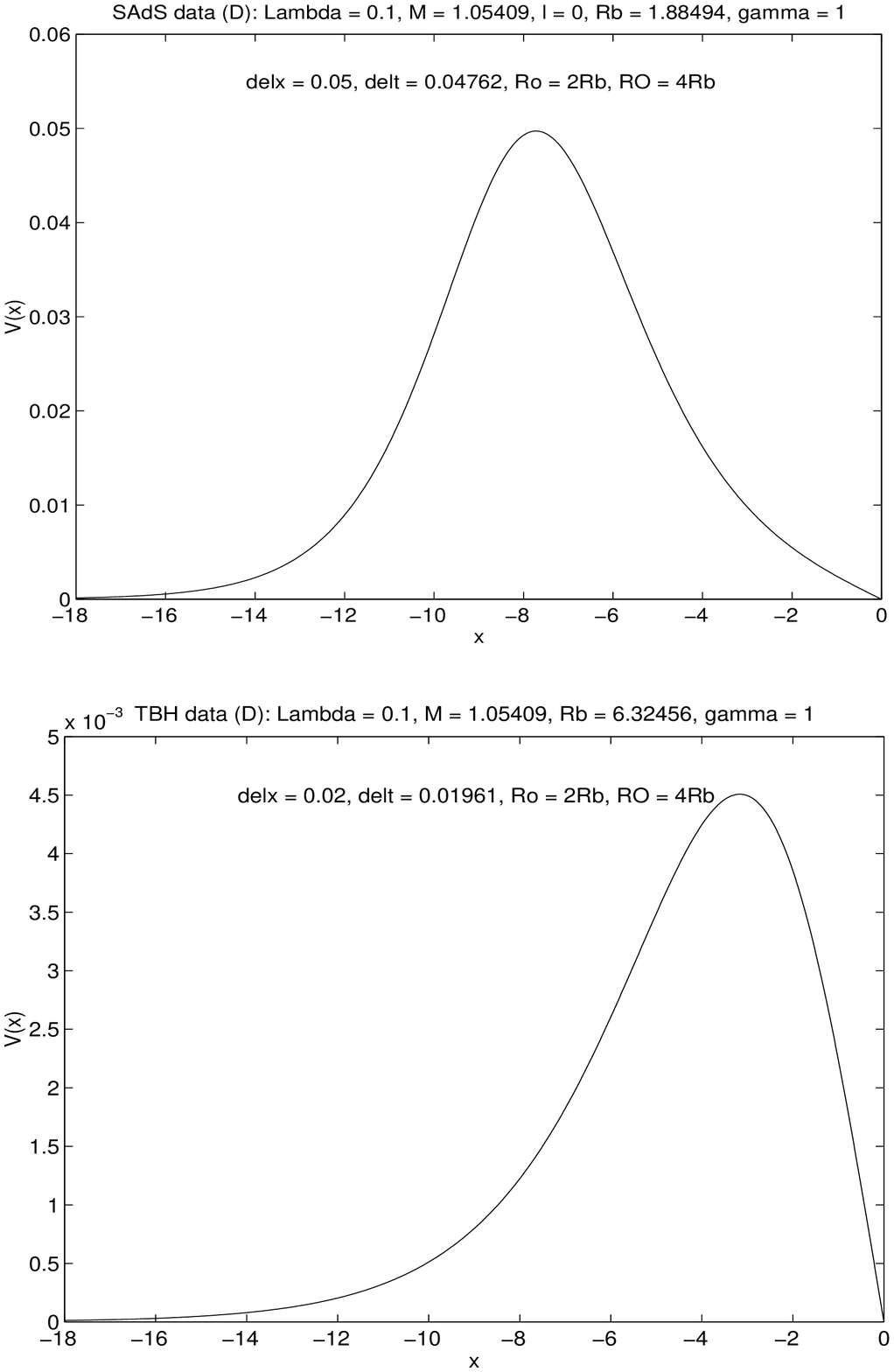,height=15cm} \hfill \mbox{}
    \caption{A comparison of the potential ${\cal V}(x)$ for the SAdS (top)
and TBH (bottom) cases.}
    \label{man-F8}
  \end{figure}

Turning next to the (neutral) topological black hole (TBH) case, 
figure \ref{man-F8}
compares the potential ${\cal V}(x)$ for $l=0$ for the SAdS (top) and TBH
(bottom) cases.  The event horizon is at $x=-\infty$, and spatial
infinity is at $x=0$. Each potential 
has a maximum at some finite $x$ and decays exponentially toward the horizon.
However the TBH potential does not have a point of inflection on the
rightward side of the maximum, unlike the SAdS case. 
Note that even if $M$ and $\Lambda$ for a given topological black hole are
equal to that of a given SAdS black hole, the location of the event horizons
$Rb$ will differ because  $V$[TBH] = $V$[SAdS]$-2$. One is then faced with
the problem of how to meaningfully
compare the falloff behaviour of a given TBH with that
of a ``similar'' SAdS black hole for differing values of $M$ and $\Lambda$.
This can be overcome by choosing to compare the SAdS and TBH cases for
equal values of the parameter $\gamma \equiv 3\sqrt{3}M/\ell = 3M\sqrt{|\Lambda|}$,
which is a dimensionless measure of the black hole mass.
This parameter governs the location of the event horizon $Rb=r_h\ell$, where
\begin{equation}\label{man-56}
r^3_h - (1-\delta_{g,1}-2\delta_{g,0})r_h 
= \frac{2}{3\sqrt{3}}\gamma
\end{equation}
defines $r_h$, for a genus $g$ black hole.
For the SAdS case ($g=0$), $0 < \gamma <\infty$,
whereas for the TBH case ($g\geq 2$) $ -1 < \gamma < \infty$ 
because subextremal negative mass black holes can also be included.
Here only the range $0 < \gamma < \infty$ will be considered.
For a given value of $\gamma$,  I shall
compare the SAdS, TBH ($g\geq 2$) and toroidal ($g=1$) 
cases over a range of values of $\Lambda$. 

\begin{figure}[htbp]
    \hfill \psfig{file=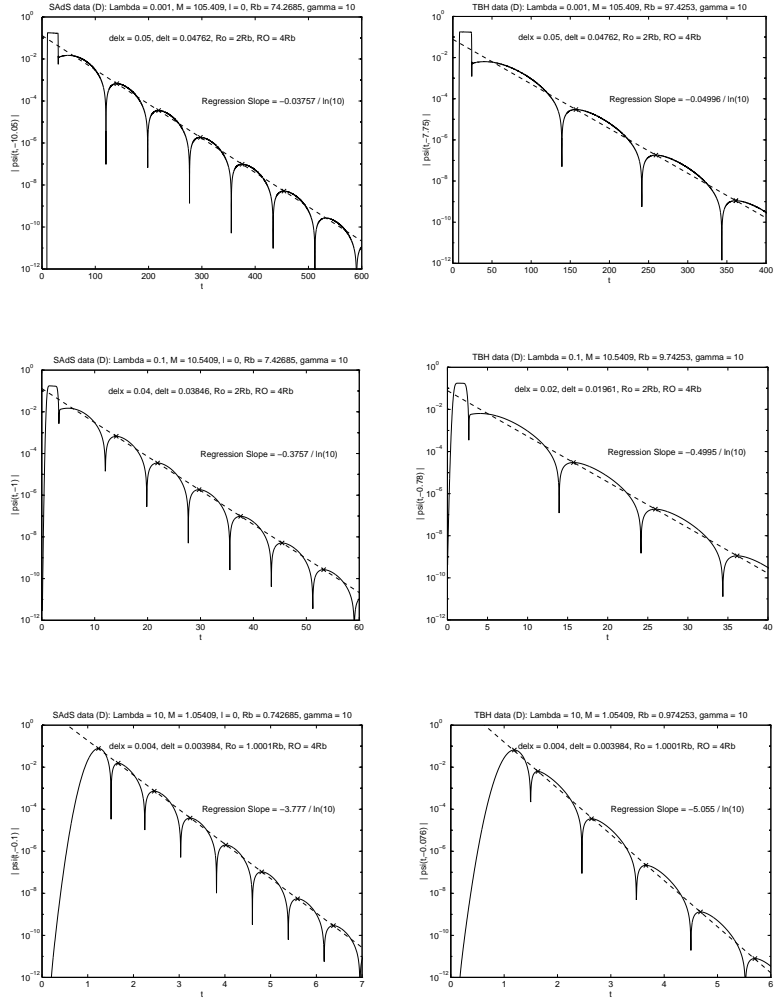,height=15cm} \hfill \mbox{}
    \caption{Falloff rates for the SAdS (left) and TBH (right) cases for $\gamma=10$
for a range of values of $|\Lambda|$.}
    \label{man-F9}
  \end{figure}

\begin{figure}[htbp]
    \hfill \psfig{file=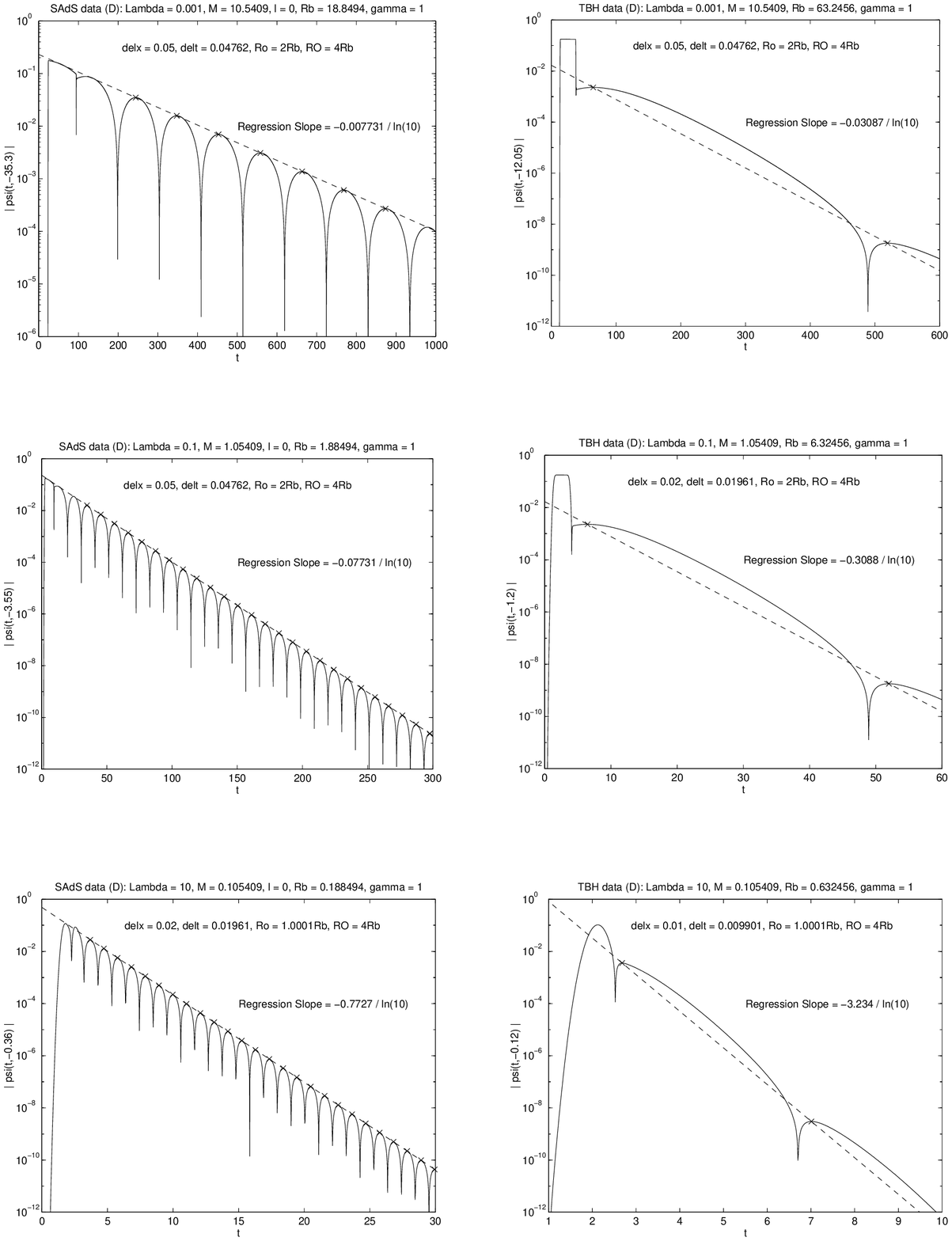,height=15cm} \hfill \mbox{}
    \caption{Falloff rates for the SAdS (left) and TBH (right) cases for $\gamma=1$
for a range of values of $|\Lambda|$.}
    \label{man-F10}
  \end{figure}

\begin{figure}[htbp]
    \hfill \psfig{file=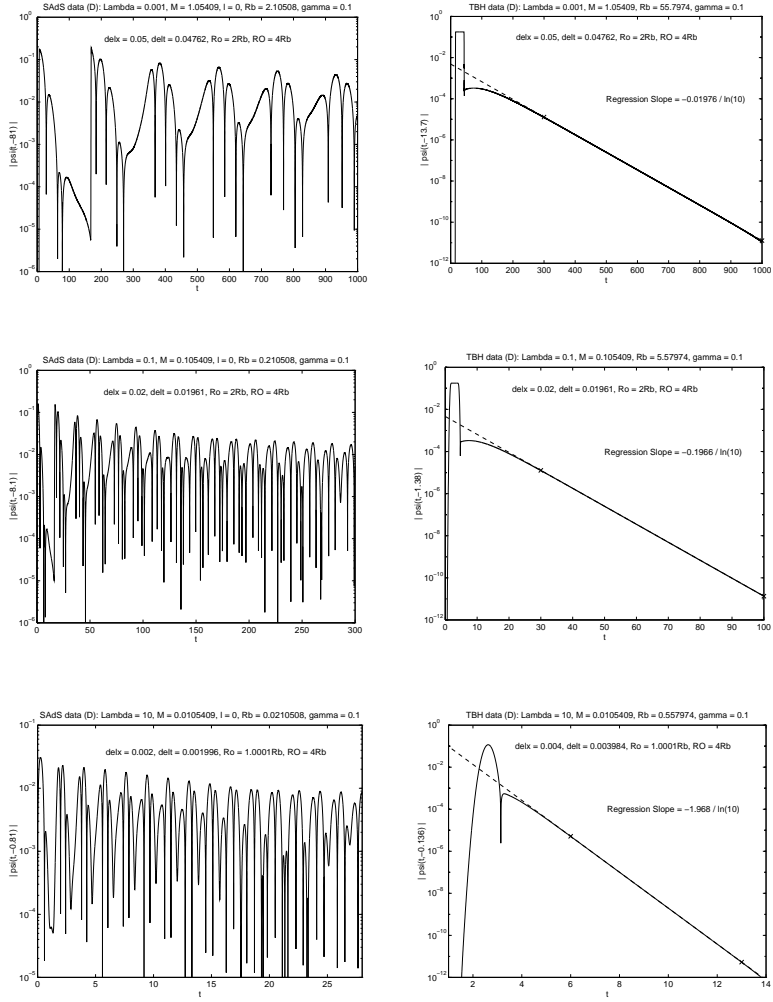,height=15cm} \hfill \mbox{}
    \caption{Falloff rates for the SAdS (left) and TBH (right) cases for $\gamma=0.1$
for a range of values of $|\Lambda|$.}
    \label{man-F11}
  \end{figure}

Consider first the cases where the genus $g\geq 2$. These cases all
have the same qualitative behaviour, shown in figures \ref{man-F9} --
\ref{man-F11}
For large $\gamma$, the  behaviour of the TBH and SAdS cases is
qualitatively similar over a wide range of values of $|\Lambda|$ with some
quantitative differences as shown in figure \ref{man-F9}.  Each exhibits a
ringing effect which falls off exponentially with time. The ringing frequency
is slightly higher and the falloff rate slightly weaker in the SAdS case that
in the TBH case.  For each, as $|\Lambda|$ increases the ringing frequency
and the falloff rate both increase.

These differences become more pronounced as $\gamma$ decreases.  From figure
\ref{man-F10}, in which $\gamma=1$, it is clear that the approximate exponential 
falloff rate in the TBH case increases substantially relative to the SAdS case. 
In addition the ringing frequency in the SAdS case rises relative to that
of the TBH case, although not as dramatically as it might first appear due
to the difference in scale on the $t$-axis of the plots. These effects become
slightly more pronounced as $|\Lambda|$ increases.  

For very small $\gamma$ substantial qualitative differences between the two cases 
emerge.  The sequence of ringing, power-law falloff and resurgence of the wave
noted above from figure \ref{man-F6} for the SAdS case returns. As $|\Lambda|$ increases
the intermediate power-law falloff is obliterated at late times, replaced by a
slowly decaying set of oscillatory ringing behaviour.  However in the TBH case
the ringing virtually disappears, being replaced by a smooth exponential falloff.
Due to limitations in computer memory and running time, it is not possible to 
tell if the ringing effect has actually vanished or if the frequency is just extremely
low, although the upper right graph in figure \ref{man-F11} would suggest that it
has actually vanished.  The falloff rate grows with increasing $|\Lambda|$.

\begin{figure}[htbp]
    \hfill \psfig{file=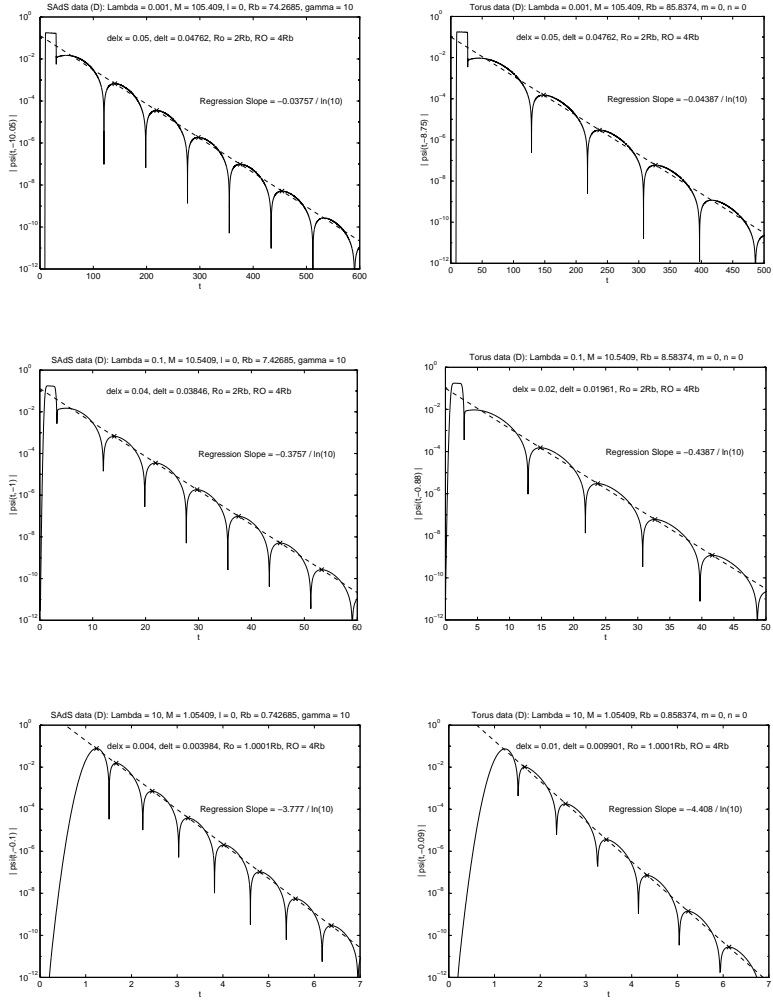,height=15cm} \hfill \mbox{}
    \caption{Falloff rates for the SAdS (left) and Toroidal (right) cases for $\gamma=10$
for a range of values of $|\Lambda|$.}
    \label{man-F12}
  \end{figure}

\begin{figure}[htbp]
    \hfill \psfig{file=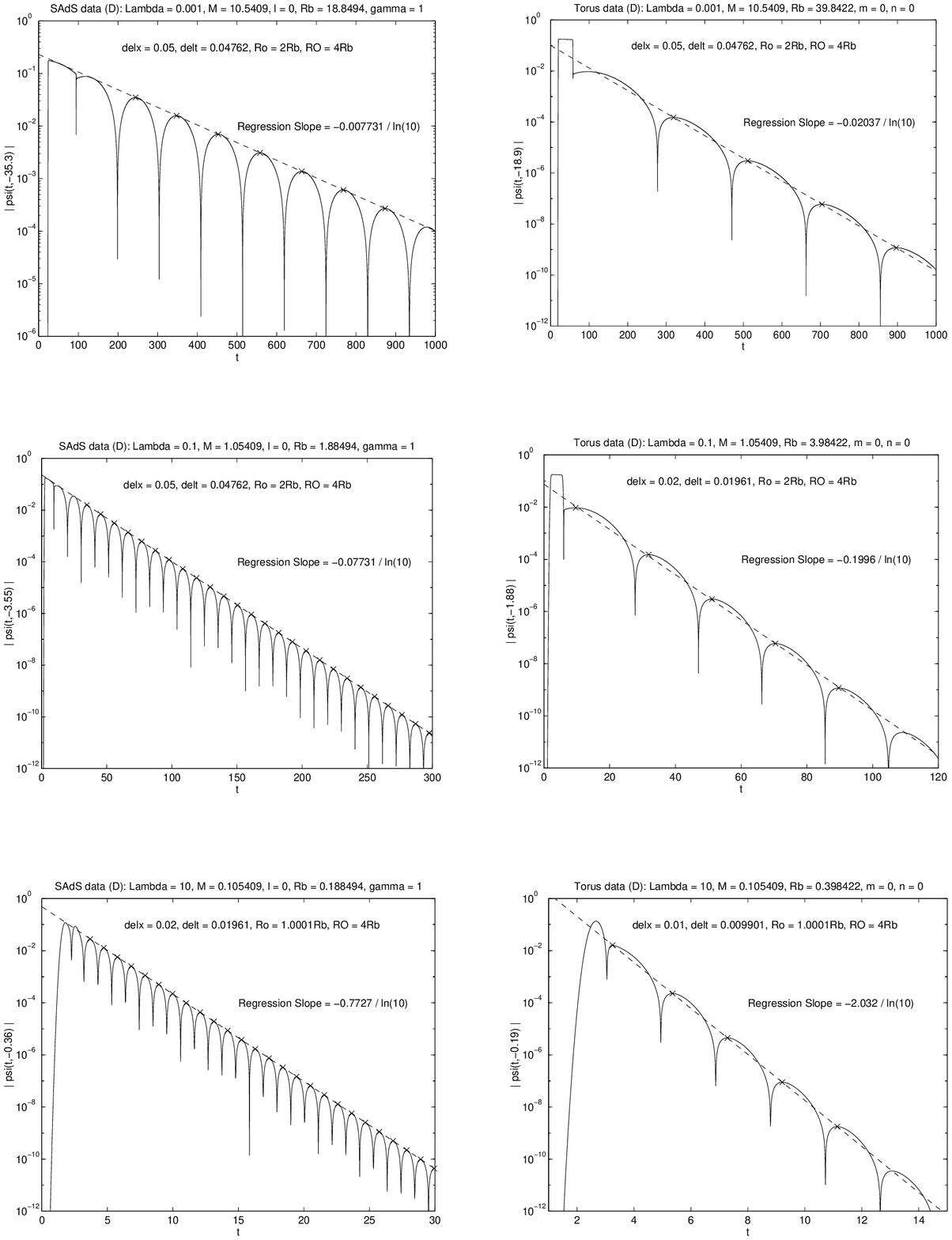,height=15cm} \hfill \mbox{}
    \caption{Falloff rates for the SAdS (left) and Toroidal (right) cases for $\gamma=1$
for a range of values of $|\Lambda|$.}
    \label{man-F13}
  \end{figure}

\begin{figure}[htbp]
    \hfill \psfig{file=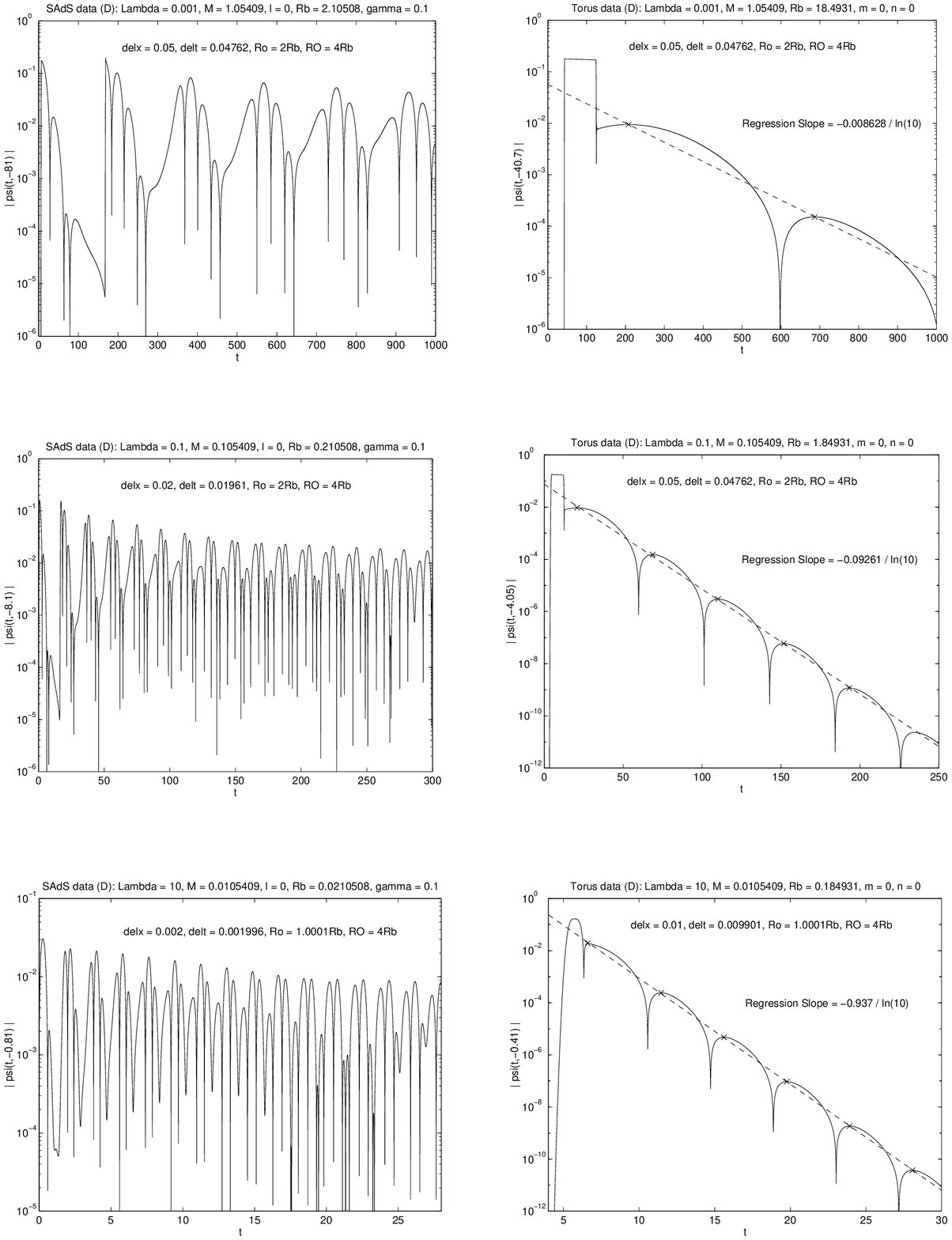,height=15cm} \hfill \mbox{}
    \caption{Falloff rates for the SAdS (left) and Toroidal (right) cases for $\gamma=0.1$
for a range of values of $|\Lambda|$.}
    \label{man-F14}
  \end{figure}

\begin{figure}[htbp]
    \hfill \psfig{file=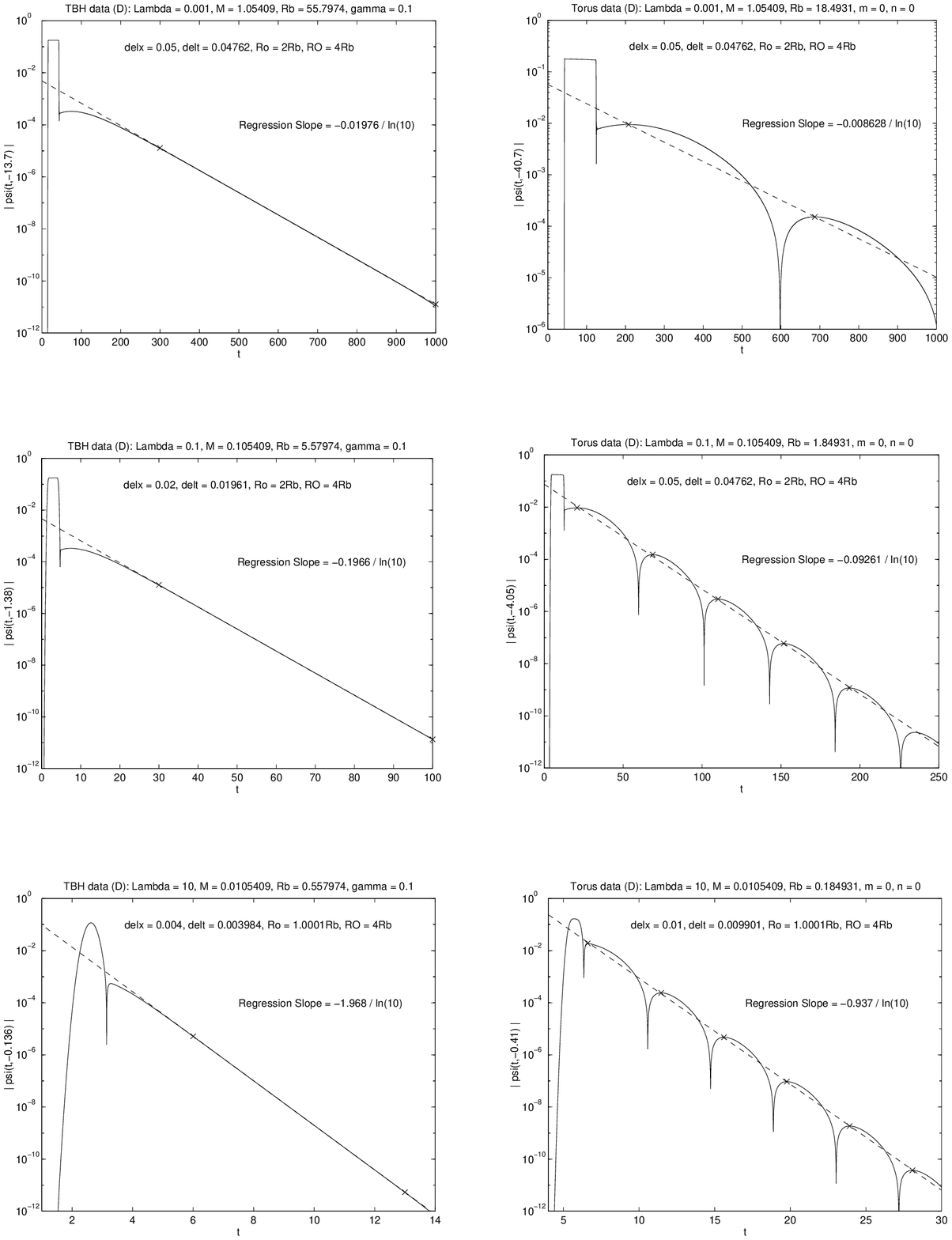,height=15cm} \hfill \mbox{}
    \caption{Falloff rates for the higher genus TBH (left) and Toroidal (right) cases 
for $\gamma=0.1$ for a range of values of $|\Lambda|$.}
    \label{man-F15}
  \end{figure}

A comparsion of the SAdS and toroidal (genus $g=1$) cases is given in figures
\ref{man-F12} -- \ref{man-F14}.  For large $\gamma$ the toroidal case is similar
to that of the genus $g\geq 2$ cases, with a slightly less rapid falloff 
and more frequent ringing for the toroidal case.  The difference is most pronounced
for small $\gamma$, with the toroidal case still exhibiting some ringing for
$\gamma = 0.1$. For clarity figure \ref{man-F15} shows a comparison at $\gamma=1$
between the higher genus (TBH) and toroidal cases.

For all of the above results, the radiative falloff behaviour for topological black holes 
is relatively insensitive to the use of Dirichlet or Neumann boundary conditions.

\section{Summary}

The radiative falloff behaviour described in the previous section indicates that
(charged or negative mass) topological black holes can indeed undergo mass inflation, 
provided the exponential falloff rate is sufficiently small.  In this case the blueshift
effect at the Cauchy horizon will overwhelm the exponential decay of the
incoming radiation, triggering mass inflation.  However it is quite conceivable that
that the falloff rate could be so strong as to cut off the mass inflation process
for some range of the parameter set $(\gamma,|\Lambda|)$. 

Whether or not this can take place is presently under investigation.
However a similar phenomenon has been observed in $(2+1)$ dimensions for the
BTZ black hole. As noted above, for $|\Lambda| J^2/M^2 > .64$, the
exponential falloff rate outside a BTZ black hole is so large that mass inflation 
is cut off.  A detailed study of the nature of the transition at this point
is presently being carried out.

Topological black holes present us with an interesting new set of possibilites to
explore in our quest to understand the physics of black holes and the role they
play in quantum gravity.  Although astrophysical applications of topological
black holes are not immediately apparent, they will necessarily play some role
in any theory of quantum gravity which includes topology changing processes.

\section*{Acknowledgements}

This work was supported by the Natural Sciences and Engineering Research 
Council of Canada. I would like to thank J.S.F. Chan, J. Creighton, N. Kaloper and
S. Solodukhin for interesting discussions on various aspects of this work.
I am particularly grateful to J.S.F. Chan who developed the
code for obtaining the results described in section V.  I would also
like to thank L. Burko and A. Ori for their kind hospitality at the
Technion Centre where these results were presented.

\end{document}